\documentclass[a4paper,fleqn]{cas-dc}

\usepackage[numbers]{natbib}
\usepackage{siunitx}

\def\tsc#1{\csdef{#1}{\textsc{\lowercase{#1}}\xspace}}
\tsc{WGM}
\tsc{QE}
\tsc{EP}
\tsc{PMS}
\tsc{BEC}
\tsc{DE}

\begin{document}
\let\WriteBookmarks\relax
\def\floatpagepagefraction{1}
\def\textpagefraction{.001}
\shorttitle{Magnetic domain wall motion in SrRuO$_3$ thin films}
\shortauthors{M Zahradn\'{i}k et~al.}

\title [mode = title]{Magnetic domain wall motion in SrRuO$_3$ thin films}                      

%

\author[1,2]{Martin Zahradn\'{i}k}[orcid=0000-0001-7660-5055]
\cormark[1]

\credit{Investigation, Formal Analysis, Visualization, Data Curation, Writing \-- Original Draft}

\address[1]{Charles University, Faculty of Mathematics and Physics, Ke Karlovu 3, 12116 Prague 2, Czech Republic}
\address[2]{Centre for Nanoscience and Nanotechnology (C2N), CNRS UMR 9001, Univ Paris-Sud, Universit\'{e} Paris-Saclay, 91120 Palaiseau, France}

\author[1]{Kl\'{a}ra Uhl\'{i}\v{r}ov\'{a}}

\credit{Conceptualization, Methodology, Investigation, Validation, Project Administration, Writing -- Review \& Editing}

\author[2]{Thomas Maroutian}

\credit{Resources, Methodology, Conceptualization, Investigation, Writing \-- Review \& Editing}


\author[2]{Georg Kurij}

\credit{Investigation}

\author[2]{Guillaume Agnus}

\credit{Supervision}

\author[1]{Martin Veis}

\credit{Conceptualization, Funding Acquisition, Writing \-- Review \& Editing, Supervision}

\author[2]{Philippe Lecoeur}

\credit{Funding Acquisition, Supervision}


\cortext[cor1]{Corresponding author}
%

\begin{abstract}
Influence of~substrate miscut on~magnetization dynamics in~SrRuO$_3$ (SRO) thin films was studied. Two films were grown on~SrTiO$_3$ substrates with high ($\sim1^{\circ}$) and low ($\sim0.1^{\circ}$) miscut angles, respectively. As expected, high miscut angle leads to~suppression of~multi-variant growth. By~means of~SQUID magnetometry, comparable relaxation effects were observed in~both the multi-variant and the nearly single-variant sample. Differences in~the magnetization reversal process were revealed by~magnetic force microscopy. It showed that the multi-variant growth leads to~higher density of~defects acting as pinning or~nucleation sites for magnetic domains, which consequently results in~deterioration of~magnetic properties. It was demonstrated that the use of~high miscut substrate is important for fabrication of~high quality SRO thin films with low density of~crystallographic defects and excellent magnetic properties.

%
\end{abstract}

\begin{graphicalabstract}
\includegraphics[scale=0.44]{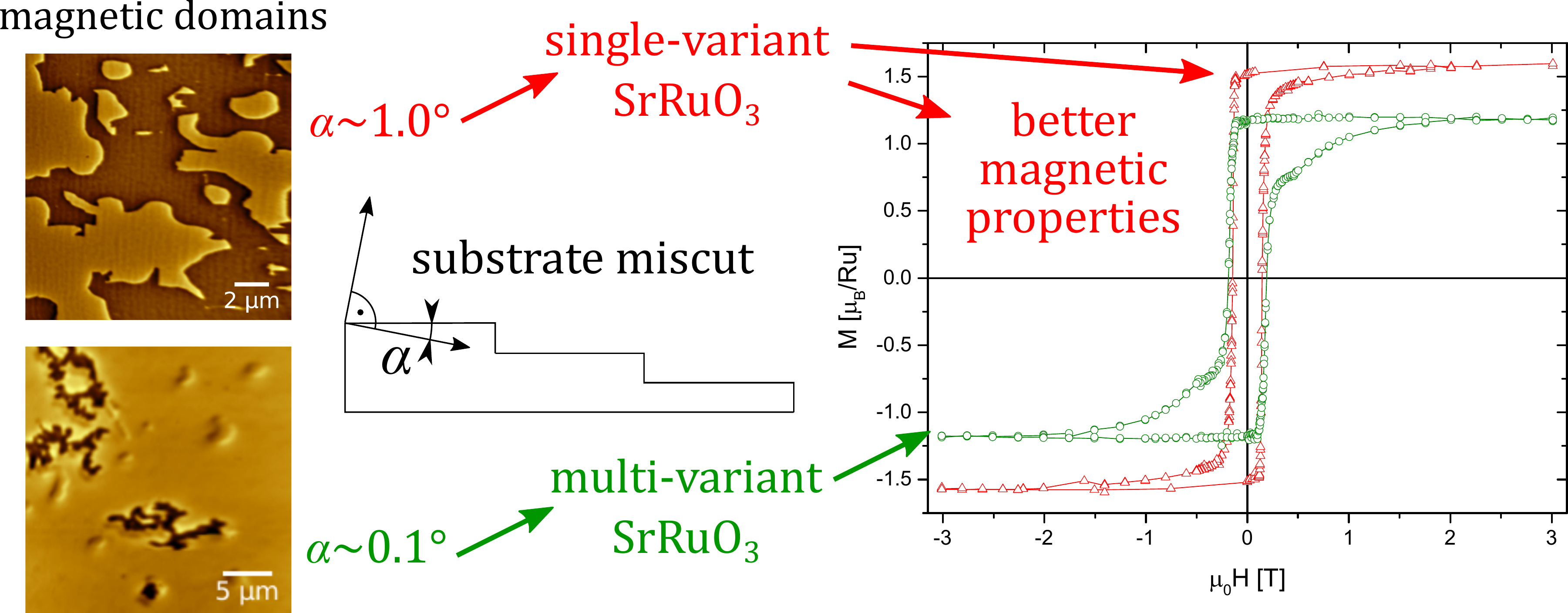}
\end{graphicalabstract}

\begin{highlights}
\item Spatially resolved dynamics of~SrRuO$_3$ magnetization was captured at~low temperatures.
\item Low-miscut angle ($\sim0.1^{\circ}$) of~SrTiO$_3$ substrate leads to~increased density of~domain nucleation and pinning centers in~SrRuO$_3$ films.
\item Presence of~anti-phase boundaries results in~magnetic structures persisting up~to~high magnetic fields ($\sim14$~T).
\end{highlights}


\maketitle

\section{Introduction}

SrRuO$_3$ (SRO) is a well known itinerant ferromagnet (T$_{C,bulk}\sim$160~K \cite{Kanbayasi1976}) that offers a broad range of~applications in~oxide electronics. Good conducting properties in~combination with nearly ideal epitaxial growth have made it the most popular material for electrode fabrication in~oxide heterostructures~\cite{Koster2012,Allouche2016}. In~addition, multilayer systems of~SRO and other oxide materials, such as SrTiO$_3$ (STO) \cite{Herranz2003} or \linebreak La$_{2/3}$Sr$_{1/3}$MnO$_3$ (LSMO) \cite{Worledge2000,Takahashi2003}, exhibit suitable properties for fabrication of~all-oxide magnetic tunnel junctions (MTJ). 

One of~the key functional elements of~the MTJ is a stable pinning layer. To~obtain a stable pinning layer, various approaches can be used. One of~the promising ways is the use of~LSMO/SRO bilayer, which exhibits antiferromagnetic coupling at~the interface, although both individual materials are ferromagnetic. While this antiferromagnetic coupling has been thoroughly investigated \cite{Ke2004,Ke2005,Padhan2006,Solignac2012}, its origin still remains unclear. To~elucidate possible mechanisms laying behind this phenomenon, at~first, detailed understanding of~magnetic behaviour of~SRO is essential. 

Despite several decades of~investigation, knowledge of~the exact nature of~magnetic anisotropy of~SRO is still lacking in~both bulk SRO~\cite{Koster2012} as well as in~thin films~\cite{Kurij2016}. Thin films of~SRO exhibit ferromagnetic ordering below T$_{C,film}\sim$150~K \cite{Eom1992} and unusual uniaxial magnetocrystalline anisotropy. \linebreak Above T$_C$ the easy axis of~magnetization lies in~the (001) orthorhombic plane and it is identical with the $b$ orthorhombic axis, i.e. its direction is $\sim$45$^{\circ}$ inclined to~the surface normal \cite{Kats2005}. Below T$_C$, the easy axis remains in~the (001) plane, however it rotates from the surface normal from $\sim$45$^{\circ}$ to~$\sim$30$^{\circ}$ with decreasing temperature \cite{Klein1996}.

This peculiar temperature dependence of~magnetocrystalline anistropy is still subject of~a scientific debate. Growth of~SRO on~the most commonly used STO substrate is possible in six different crystallographic orientations \cite{Jiang1998a}. However, it has been found that the anisotropy is independent of~the orientation \cite{Marshall1999}. Others attribute changes in~the magnetocrystalline anistropy to~distortions of~the SRO unit cell during the deposition process \cite{Gan1999}. Kolesnik \textit{et al.} \cite{Kolesnik2006} argue the effect of~twinning, comparing the anisotropy in twinned and untwinned SRO films.

An~additional important issue is arising as attempts of magnetization switching in~SRO films are emerging. After current induced domain wall nucleation \cite{Feigenson2008} and domain wall motion~\cite{Feigenson2007} was presented, temperature induced \cite{Sarkar2013} as well as current induced \cite{Shperber2012,Shperber2013} magnetization reversal in~SRO films was demonstrated. Another study reported on~periodic control of~magnetization via piezoelectric substrate in~SRO/Pb(Mg$_{1/3}$Nb$_{2/3}$)O$_3$-PbTiO$_3$ heterostructure \cite{Zhou2014}. All such attempts lead to~possible applications in~spintronic devices. For their proper functioning, not only a precise description of~magnetic anisotropy, but also detailed knowledge of~dynamic behaviour of~magnetic domains is required.

Magnetic domain dynamics consists of~two main mechanisms: domain nucleation, and domain wall motion (propagation). It has been already observed by~Barkhausen in~1919, when he detected the so called Barkhausen noise, that the magnetization reversal process is not continuous~\cite{Barkhausen1919}. It is due to~the fact that both formation of~domains, and domain wall motion needs activation energy to~overcome a critical domain size, and to~release the domain walls from pinning centers, respectively. Visualisation of~the magnetization reversal can be realized not only by~measuring Barkhausen noise, but also by~direct techniques. Especially when the dynamics is slow enough, magnetic domains can be observed, e.g. by~Kerr microscopy~\cite{Pommier1990}, or~even magnetic force microscopy (MFM)~\cite{Schwarz2004}.

This study reports on~time evolution of~magnetic domains in~multi-variant and nearly single-variant SRO thin films. Superconducting quantum interference device (SQUID) magnetometry and magnetic force microscopy (MFM) was used to~investigate the magnetization dynamics and magnetic domain formation of~both the multi-variant and the nearly single-variant SRO thin films. Pronounced differences in~the magnetic domain wall motion behaviour were observed. It was argued that those differences originate in~crystallographic defects induced by~multi-variant growth of~the films. Such findings are of~high importance for design and realisation of~all-oxide MTJ and new spintronic devices, because desired magnetic properties in~SRO films could be tuned either by~proper selection of~substrate miscut angle, or~even more generally by~any parameters of~the deposition process that directly influence crystallinity of~the films.

Moreover, the present study demonstrates a pioneering aspect in~investigation of~SRO magnetic properties, which can be further generalised even for other material systems with low Curie temperature. Proper investigation of~such materials ($T_C<$ room temperature) in~terms of~magnetic properties is a challenging task requiring complex experimental techniques. So far, a successful observation of~magnetic domains in~SRO thin films was demonstrated by~means of~Lorentz transmission electron microscopy~\cite{Marshall1999}. This technique, however, puts additional requirements for sample pre\-paration, which leads to~further difficulties during the measurement process itself, making the whole experimental procedure more complex and difficult. On~the other hand, the low temperature MFM is powerful enough to~provide access to~information about~magnetic properties while putting none additional requirements on~sample preparation, making the experimental procedure easily feasible. There has already been one MFM study observing magnetic domains in~SRO~\cite{Landau2012}, however, the SRO was in~a form of~patterned nanoislands, and therefore exhibiting different magnetization dynamics compared to~this study. Here the first MFM study of~magnetization reversal in~SRO thin films is presented, demonstrating the potential of~this method in~complex investigation of~ferromagnetic materials with low Curie temperature.

\section{Experimental details}

Investigated SRO films were prepared by pulsed deposition on~(001) oriented STO substrates with Ti termination. Growth of~SRO on~STO substrate is possible in~six different crystallographic orientations, so called variants~\cite{Jiang1998a,Jiang1998b}. Low miscut angle of~the substrate leads to~coexistence of~several variants, i.e. to~growth of~polycrystalline films. Higher miscut angle leads to~suppression of~multi-variant growth~\cite{Gan1997,Jiang1998a,Jiang1998b}, therefore we used substrates of~1$^{\circ}$ and 0.1$^{\circ}$ of~miscut angles to~achieve growth of~single-variant (SRO1) and multi-variant (SRO2) films, respectively.

The pulsed laser deposition process was carried out under background oxygen pressure of~120~mTorr. A KrF laser at~a wavelength of~248~nm was used, with typical growth rate of~15 pulses per monolayer and 2~Hz pulse-repetition rate. The STO substrate was kept at~900~K during the deposition process. Such parameters lead to~single or~poly-crystalline growth of~SRO, depending on~miscut angle of~the STO substrate. 

Proper crystallinity and surface morphology of~the SRO films was verified by~X~ray diffraction (XRD) and atomic force microscopy (AFM), respectively. The XRD analysis was carried out using a PANanalytical X'Pert PRO diffractometer, measuring reciprocal space maps (RSM) around (204) family of~STO Bragg reflections. SRO1 was found to~be nearly single-variant, while presence of~two crystallographic twins was revealed in~SRO2. Thicknesses of~the films were determined by~the same instrument, symmetric scans around (002) Bragg reflection of~SRO fitted by~classical interference formula gave values of~28~nm (SRO1) and 46~nm (SRO2), respectively. The AFM images were taken at~room temperature by~a Bruker Dimension Edge AFM microscope.

The magnetization process was measured by~SQUID magnetometer (Quantum design, MPMS7 XL, RSO option). To observe the change of~magnetic domains in~SRO thin films, low temperature atomic force microscope attoAFM/MFM Ixs was used, inserted in~PPMS~14 (cryostat). The PPMS 14 does not allow a real zero-field cooling due to~residual field of~the superconducting coil, so the focus was made only on~the magnetization process from fully magnetized state. The samples were magnetized at~magnetic field of~+3~T or higher (a control MFM scan was performed back at~zero field for each sample). After that two measurement modes were chosen: (i) small negative field was applied and kept during the MFM measurement, (ii) small negative field was applied, and the MFM measurement was then performed at~zero field to~avoid further magnetization reversal process. MFM data were analyzed and plotted using Gwyddion software~\cite{Necas2012}.

\section{Results and discussion}

\subsection{Crystallographic properties and morphology}

\begin{figure*}
	\centering
	\includegraphics[scale=0.51]{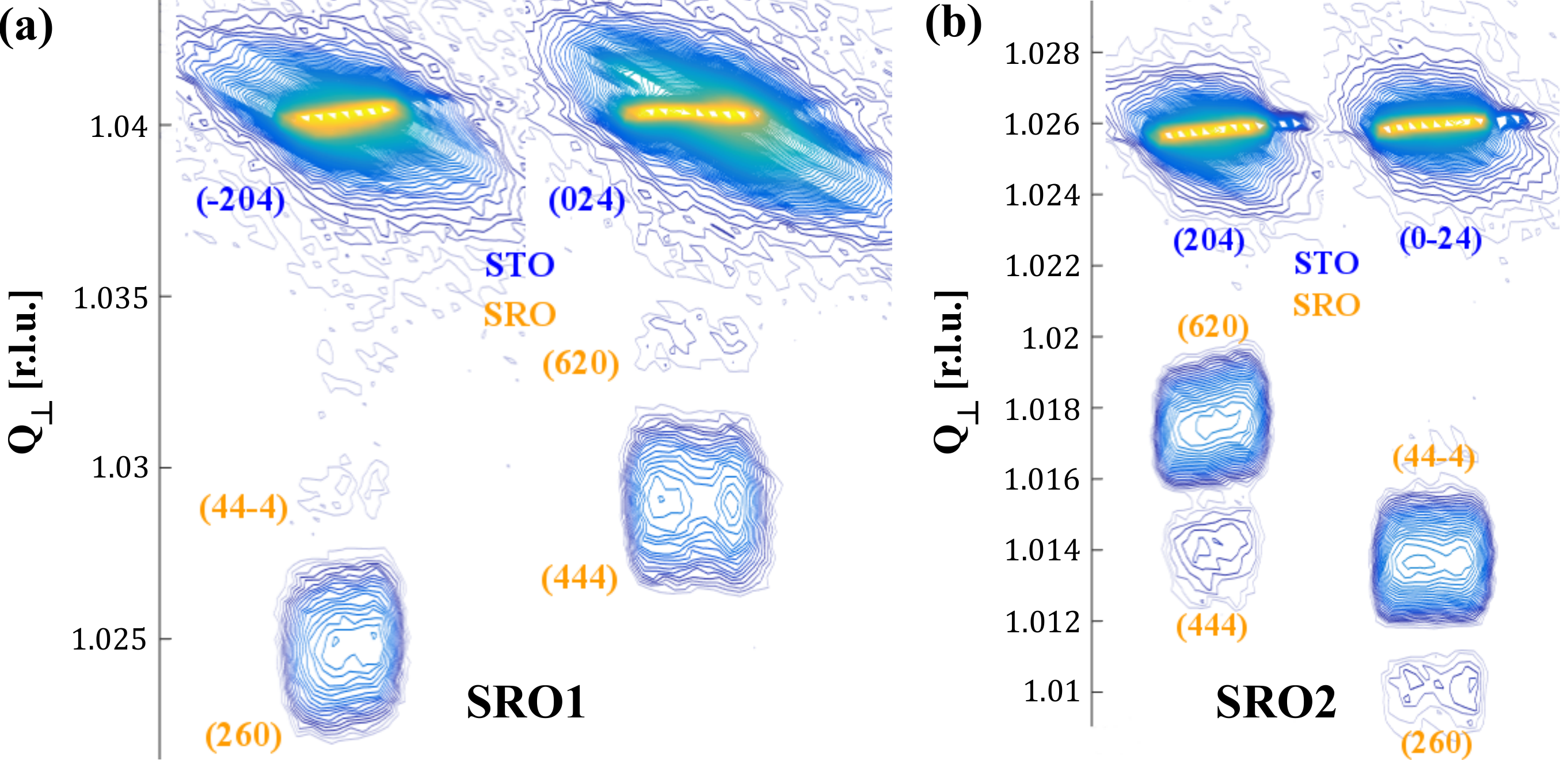}
	\caption{Reciprocal space maps around (204) family of~Bragg reflections of~SrTiO$_3$ measured on~SrRuO$_3$ thin films deposited on~(a) vicinal SrTiO$_3$ substrate (SRO1), ratio of~the SrRuO$_3$ peaks was estimated as 1:9, i.e. the film is nearly single-variant; (b) SrTiO$_3$ substrate of~low miscut angle (SRO2). Presence of~two crystallographic SrRuO$_3$ twins is clearly visble, their ratio was estimated to~be approximately 1:2.}
	\label{fig:SRO_RSM}
\end{figure*}

In~order to~determine crystallinity of~the samples, RSM were measured for both SRO1 and SRO2. The results are shown in~Figs.~\ref{fig:SRO_RSM}(a) and (b), respectively. The measurements were carried out at~two azimuths for each sample. The out-of-plane component of~reciprocal lattice vector $Q_{\perp}$ is given as~the out-of-plane projection of~$|Q|=2\sin(\theta)/\lambda$, where $\theta$ is the Bragg diffraction angle and $\lambda=1.5406$~\AA.

Fig.~\ref{fig:SRO_RSM}(a) shows the RSM of~SRO on~vicinal STO substrate (high miscut angle). The same lateral position of~both STO and SRO peaks indicates that the film remains fully strained. Although a growth of~only one crystallographic variant was expected on~vicinal STO substrate, the RSM show presence of~a second SRO variant as well. Above the main SRO peak a blurred side peak is visible. From intensity maxima the ratio of~SRO peaks was roughly estimated as 1:9, which shows that the fraction of~second crystallographic variant is very low, i.e. the SRO1 film is nearly single-variant.

Fig.~\ref{fig:SRO_RSM}(b) shows the RSM of SRO2. They reveal fully strained SRO even at higher thickness of~the deposited layer. A growth of~the second crystallographic variant was expected, and indeed the SRO side peaks are clearly visible. The volume of~the second variant is significantly higher compared to~SRO1. For SRO2 the ratio of~SRO peaks was estimated to~approximately 1:2. This means that the second crystallographic variant represents around 30\% in~case of~SRO2, while it represents only 10\% in~case of~SRO1.

A brief note should be made at~this point, emphasizing that the thickness variation between the two films is not a leading cause of~appearance of~the second crystallographic variant. As already demonstrated by~several research groups, the key factors in~determining the crystallinity of~SRO are the miscut angle and step direction of~the STO substrate~\cite{Jiang1998a,Jiang1998b,Gan1997,Vailionis2007}. Multiple variants could appear due to~partial relaxation at~higher thicknesses of~the SRO layer, however, as demonstrated in~Fig.~\ref{fig:SRO_RSM}, both investigated films are fully strained, leaving the substrate miscut angle as driving parameter for the observed differences in~crystallinity.

\begin{figure*}
	\centering
	\includegraphics[scale=0.8]{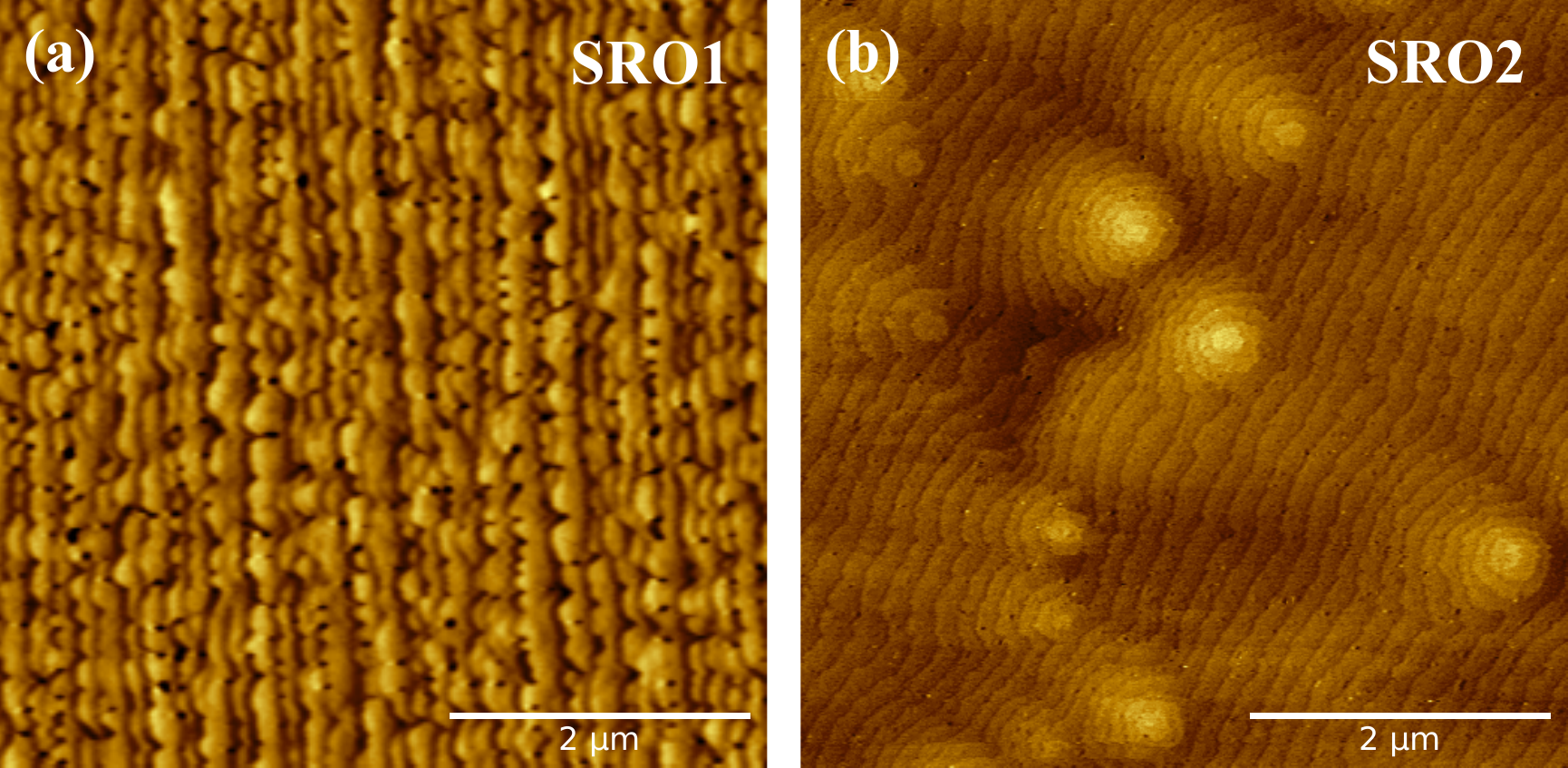}
	\caption{AFM images (5$\times$5~\si{\micro\metre\squared}) of~(a) 28~nm thick nearly single-variant SrRuO$3$ film (SRO1), surface roughness (RMS) is 1.6~nm; (b) 46~nm thick multi-variant SrRuO$_3$ film (SRO2), surface roughness (RMS) is 0.3~nm. Atomic steps are clearly visible for both films (vertical direction for SRO1, diagonal for SRO2), however the step bunching makes it a little less obvious for SRO1. For SRO2 three-dimensional islands can be seen as well.}
	\label{fig:SRO_AFM}
\end{figure*}

Both samples, the nearly single-variant SRO1 and the multi-variant SRO2 were characterized by~room temperature AFM as presented in~Figs.~\ref{fig:SRO_AFM}(a) and (b), respectively. In~both samples, atomic steps are clearly visible, which is a signature of~good epitaxial growth of~the films. Difference in~the step width corresponds well to~the different miscut angle of~the substrates. In~addition to~atomic steps, Fig.~\ref{fig:SRO_AFM}(b) shows several island-like feature, which are a clear indication of~3D growth at~higher thicknesses of~deposited SRO layer. Higher surface roughness of~SRO2 (1.6~nm) compared to~SRO1 (0.3~nm) is due to~high miscut angle of~the vicinal substrate and step bunching during SRO growth~\cite{Esteve2011}.

\subsection{Magnetic properties}

\begin{figure*}
	\centering
	\includegraphics[scale=0.9]{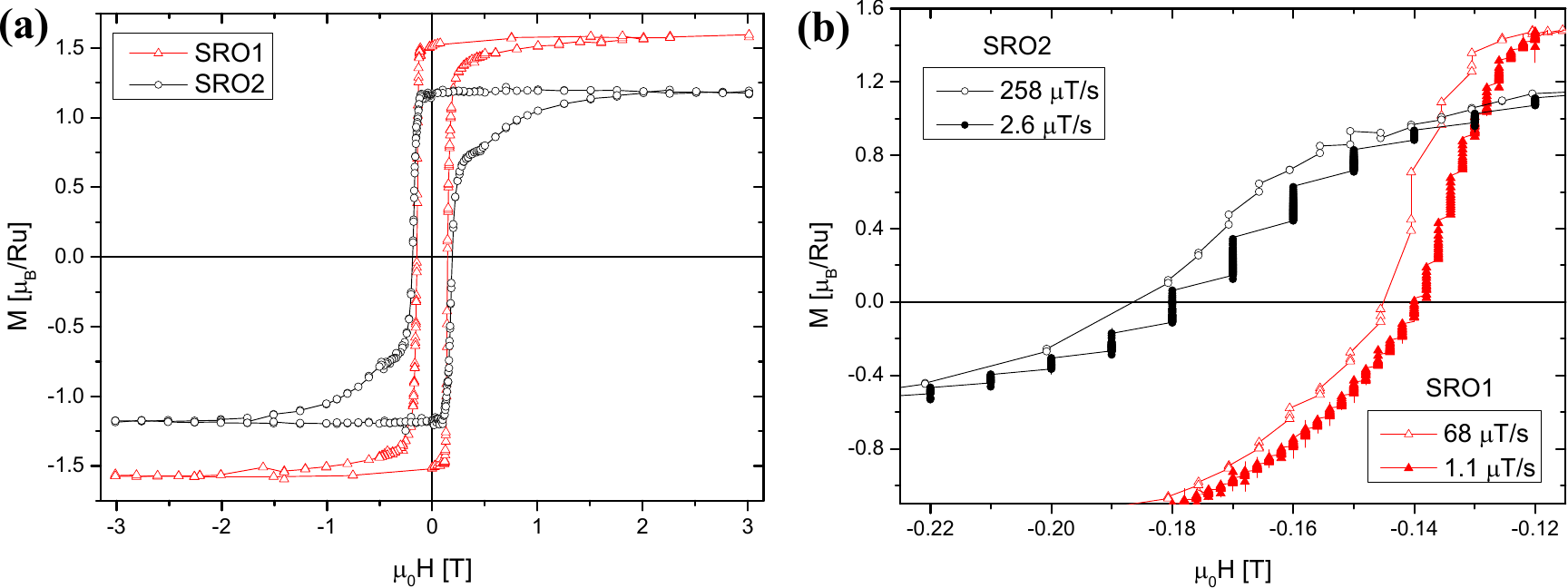}
	\caption{(a) Hysteresis loops of~magnetization, (b) zoom of~the hysteresis loops of~magnetization near coercive field, measured at~fast and slow rates. The data were recorded on~nearly single-variant (SRO1) and multi-variant (SRO2) SrRuO$_3$ films by~SQUID magnetometry at~20~K with magnetic field applied perpendicular to~the sample surface. The pronounced field steps in~(b) represent the actual field changes during the measurement. The slow ramping rates were realised by~multiple data recording for a fixed period of~time at~a given value of~the magnetic field, for which a gradual drop of~magnetization can be observed at~these values. See the text for more detailed explanation of~the experimental procedure.}
	\label{fig:SQUID}
\end{figure*}

The magnetization process was firstly studied by~magnetometry measurements. In~Fig.~\ref{fig:SQUID}(a), magnetization loops of~SRO1 and SRO2 samples are presented, measured at~$T=20$~K with external magnetic field applied perpendicular to~the sample surface. The field was ramped at~68 and \SI[per-mode=symbol]{258}{\micro\tesla\per\second}, respectively. At~these fast rates the coercive fields \linebreak $\mu_0H_{C1,fast}=145$~mT and $\mu_0H_{C2,fast}=185$~mT were determined. The measurement temperature of~20~K was chosen as optimal value with respect to~Curie temperature ($T_{C,film}\sim150$~\cite{Eom1992}) as well as with respect to~the temperature-de\-pend\-ence curve of~the SRO magnetic moment~\cite{Kurij2016}. 

Magnetic moment of~SRO in~saturation at~low temperature is 1.6~$\mu_B/$Ru~\cite{Bushmeleva2006}. Saturation magnetization determined from Fig.~\ref{fig:SQUID}(a) reaches 1.5~$\mu_B/$Ru for SRO1, demonstrating high quality of~the deposited SRO layer. The value determined for SRO2 is 1.2~$\mu_B/$Ru, which is slightly lower compared to~SRO1. This might be due to~the presence of~two SRO varians, and therefore lower crystalline quality of~the sample. Presence of~multiple crystallographic variants leads to~intermediate areads among them, where the crystalline structure is not exactly defined. However, we assume that the effect would be very small as no significant additional broadening is visible in~the XRD measurements. Unfortunately we are not aware of~any previous research investigating this topic more in~detail. Another possible explanation is discussed later in~the section of~Magnetic domains imaging.

Looking at~the shape of~the loops presented in~Fig.~\ref{fig:SQUID}(a), one can see that both loops exhibit square-like behaviour typical for loops measured along an easy axis of~magnetization, which confirms that easy axis of~both films has an out-of-plane component. One can also notice one particular difference in~behaviour of~these two loops. Around $\pm0.5$~T a small, but clearly remarkable drop of~magnetization can be observed for SRO2. This shape of~the hysteresis loop indicates two different contributions to~the magnetic moment coming from the two different crystallographic variants. For SRO2 it is visible due to~higher representation of~the second SRO variant, while the magnetization drop becomes indistinct for SRO1. It confirms that the volume of~the second variant is negligible, and that the SRO1 sample is nearly single-variant.

Fig.~\ref{fig:SQUID}(b) shows the difference in~the magnetization process when ramping the magnetic field at~different rate. For each sample there is a zoomed part of~the loop presented in~Fig.~\ref{fig:SQUID}(a) compared to~a loop measured at~a slower rate, for SRO1 it is \SI[per-mode=symbol]{1.1}{\micro\tesla\per\second}, for SRO2 it is \SI[per-mode=symbol]{2.6}{\micro\tesla\per\second}. In~case of~the slow loops, lower values of~the coercive field were found, $\mu_0H_{C1,slow}=140$~mT and $\mu_0H_{C2,slow}=180$~mT for SRO1 and SRO2, respectively. The difference represents 5~mT for both samples, clearly demonstrating a difference in~the dynamics of~the magnetization reversal process. For the slow field ramping in~Fig.~\ref{fig:SQUID}(b), multiple data points for each field value are visible. That is because in~the vicinity of~coercive field, the magnetic field was kept at~a constant value for 27 (SRO1) and 60 (SRO2) min while measuring a set of~few hundred data points. Thus for each value of~the magnetic field time evolution of~magnetization is clearly visible. Lines in~the figure are guides to~the eye, connecting the data points in~chronological order of~recording. The average ramping rate is then calculated across the whole region around coercive field.

A note should be made here addressing the issue of slight\-ly different thicknesses of~the investigated films. One might express doubts, whether any differences observed in~the magnetization behaviour could not be attributed to~changes of~this parameter. However, any thickness dependent changes in~magnetization behaviour (such as $T_C$ or~magnetic anisotropy) are typically governed by~interfacial effects, and therefore they are taking place in~the ultrathin region of~thicknesses below $\sim$9~monolayers~\cite{Koster2012,Xia2009,Ishigami2015}. Thicknesses of~the films presented here are well above this threshold for the so-called thick film behaviour, where no such effects were shown to take place. The only quantity that continues changing beyond this limit remains to~be the saturation magnetization. However, data presented in~Fig.~\ref{fig:SQUID} are corrected for these changes via normalisation to~formula units, showing that the thicker film exhibits even lower saturation magnetization, which further supports variants-related phenomena as a key parameter in~determining the magnetic properties. Magnetization changes could also take place due to~strain relaxation of~the films with increasing thickness, however as shown in~Fig.~\ref{fig:SRO_RSM}, both films remained fully strained up~to~46~nm of~film thickness, which safely rules out any potential influence of~thickness variation on~magnetic properties of~the SRO films.

\subsection{Magnetic domains imaging}

\begin{figure*}
	\centering
	\includegraphics[scale=0.65]{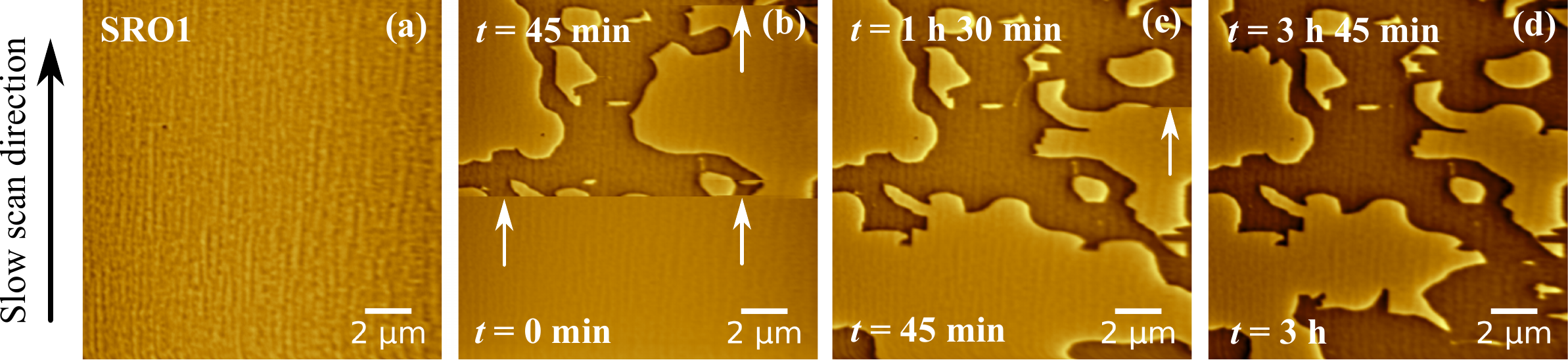}
	\caption{MFM images (15$\times$15~\si{\micro\metre\squared}) of~magnetization reversal in~nearly single-variant SrRuO$_3$ film (SRO1) measured at~20~K with field applied perpendicular to~the sample surface. Slow scan direction was vertical, proceeding upwards, as indicated by~the black arrow. (a) Fully saturated state measured at~+3~T, (b) first scan at~-119~mT, where beginning of~the switching process can be seen, (c) second scan at~-119~mT measured right after the first scan, (d) last scan at~-119~mT after nearly 4~h, further magnetization evolution is still visible. White arrows in~(b) and (c) point out horizontal division lines between bright (initial) and dark (reversed) areas that are just being switched during the scan. Areas below these lines appear switched in~following image. The time after field application is indicated for the beginning (bottom) and the end (top) of~each scan.}
	\label{fig:SRO1_MFM}
\end{figure*}

The magnetization process was visualised employing low temperature magnetic force microscopy. A typical series of~MFM scans of~SRO1, taken at~20~K, is shown in~Fig.~\ref{fig:SRO1_MFM}. Duration of~each scan was approximately 45 min. The slow scan direction is indicated by~a black arrow in~the figure. The first scan (shown in~Fig.~\ref{fig:SRO1_MFM}(a)) was taken after saturating the sample at~+3~T. One can see a homogeneously magnetized area of~the sample with a signature of~the atomic steps, coming from the crosstalk of~topography (cf.~Fig.~\ref{fig:SRO_AFM}(a)). Then a small negative field of~-119~mT was applied and kept during the measurement. The beginning of~magnetization reversal process is shown in~Fig.~\ref{fig:SRO1_MFM}(b). Bright areas represent the initial magnetization, while the dark areas are reversed. The homogeneous bottom half of~the image is still fully saturated. After approximately 20 min, the large bright area abruptly ends as the first switching event is covered by~the horizontal movement of~the scanner. As expected from the SQUID magnetization measurements (cf. Fig.~\ref{fig:SQUID}(b)), the magnetization reversal process further continues in~time as shown in~Fig.~\ref{fig:SRO1_MFM}(c) taken right after the first scan while keeping the same field of~-119~mT. Further increment of~the reversed (dark) area can still be observed after nearly 4~h at~the same magnetic field, as demonstrated in~Fig.~\ref{fig:SRO1_MFM}(d). Such slow time evolution of~the magnetic domain pattern is in~agreement with the magnetization relaxation observed by~SQUID (cf. Fig.~\ref{fig:SQUID}(b)). However, it is the contrary of~an MFM study of~magnetization reversal in~patterned nanoislands of~SRO, presented by~Landau~\textit{et al.}~\cite{Landau2012}. They observed no relaxation effects in~the SRO nanoislands. Such behaviour can be satisfactorily explained in~terms of~the strong shape anisotropy induced by~patterning into the nanostructures. On~the other hand, there is no such contribution to~magnetic anisotropy in~our films, therefore the relaxation effects remain observable.

\begin{figure*}
	\centering
	\includegraphics[scale=0.65]{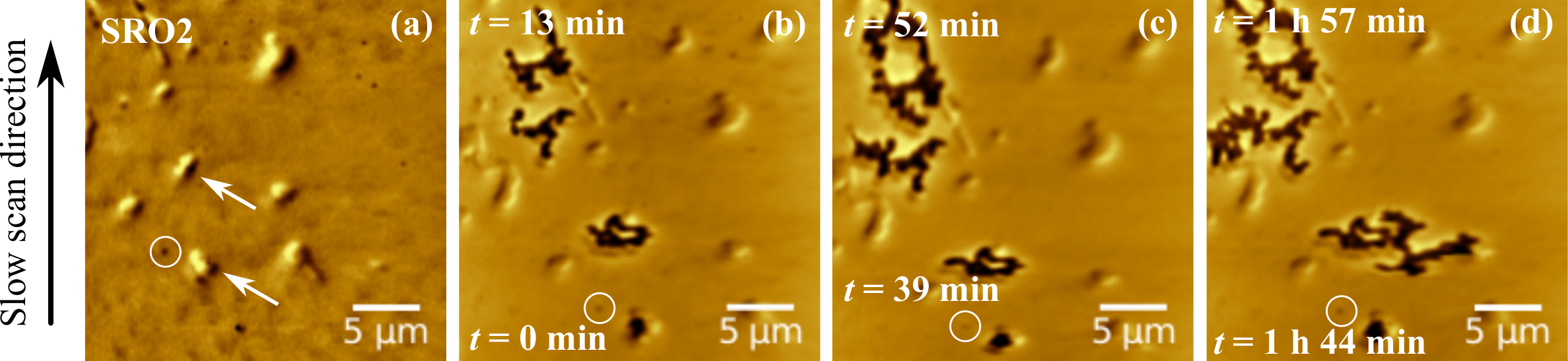}
	\caption{MFM images (26$\times$26~\si{\micro\metre\squared}) of~magnetization reversal in~multi-variant SrRuO$_3$ film (SRO2) measured at~20~K with field applied perpendicular to~the sample surface. Slow scan direction is indicated by~the black arrow. Bright areas represent initial magnetization, dark areas are reversed. (a) Fully magnetized state, measured in~remanence after saturation in~+3~T, (b) first scan at~-140~mT where domain nucleation is captured, (c) one of~the following scans at~-140~mT measured after 52 min, (d) last scan at~-140~mT after nearly 2 h. White circle in~all images highlights a dirt partcile that serves as a marker. White arrows in~(a) indicate bubble-like features acting as domain nucleation centers. The time after field application is indicated for the beginning (bottom) and the end (top) of~each scan.}
	\label{fig:SRO2_MFM}
\end{figure*}

Fig.~\ref{fig:SRO2_MFM} shows a series of~MFM scans taken at~20~K on~SRO2. Duration of~each scan was approximately 13 min. Fig.~\ref{fig:SRO2_MFM}(a) shows saturated state measured in~remanence after application of~+3~T. One can see a homogeneously magnetized area with two kind of~features. First, weak dark shadow spots, coming from the crosstalk of~topography (cf. Fig.~\ref{fig:SRO_AFM}(b)), are visible all over the investigated area. Then, several bubble-like features with non-regular shape can be seen across the image. Two of~them are indicated by~white arrows. These features clearly exhibit magnetic signal that cannot be erased even in~magnetic fields up~to~14~T, which was the largest field in~our experimental setup. The exact nature of~the bubble-like features was not unambiguously clarified, but we assume that they can be related to~crystallographic defects, such as anti-phase boundaries (APB)~\cite{Zijlstra1979}, arising in~the multi-variant growth. Such defects may consequently lead to~creation of~small areas with antiferromagnetic ordering, whose magnetic signal can persist up to~high magnetic fields, as reported for example in~magnetite~\cite{Margulies1997}. Then these crystallographic defects can also act as domain nucleation centers, which indeed was observed by~means of~MFM.

Fig.~\ref{fig:SRO2_MFM}(b) shows the MFM scan taken after application of~small negative field of~-140~mT, where the two bubble-like features indicated by~arrows in~Fig.~\ref{fig:SRO2_MFM}(a) become domain nucleation centers. The dark magnetic domains are clearly originating in~these two persistent magnetic structures. The reversed area of~the image increases in~time as shown in~Fig.~\ref{fig:SRO2_MFM}(c), taken after 52 min from the initial field application. As demonstrated in~Fig.~\ref{fig:SRO2_MFM}(d), the magnetization pattern still evolves after nearly two hours, exhibiting similar timescale of~several hours as in~case of~the magnetization relaxation in~SRO1 (cf. Fig.~\ref{fig:SRO1_MFM}). The only apparent difference in~the magnetization reversal process then remains in~the size of~the magnetic domains, which are markedly smaller in~the SRO2 sample. Note that the scan size in~Fig.~\ref{fig:SRO2_MFM} is larger compared to~scans in~Fig.~\ref{fig:SRO1_MFM}, as we were trying to~cover larger area when searching for the domain nucleation process. Yet the difference in~the domain size is evident, suggesting that the density of~pinning centers is higher in~SRO2, which leads to~more indented domain pattern. 

A note should be made here that due to~different densities of~pinning centers the relaxation effects might be expected to~occur on~different timescales in~both films. Although the magnetization relaxation was observed on~comparable timescales of~several hours, the relaxation behaviour was not investigated on~longer timescale, where such possible differences might become noticeable.

The bubble-like features also lead to~another important observation. As their magnetic signal is persistent up~to~high magnetic fields ($\sim14$~T), in~the hysteresis loop of~SRO2 measured up~to~3~T (see Fig.~\ref{fig:SQUID}(a)) there is no observable change of~the magnetization slope associated with these features. However their mere presence automaticaly leads to~decrease of~the saturated area of~the sample, which should manifest itself via decrease of~the overall saturation magnetization. That is exactly the result presented in~Fig.~\ref{fig:SQUID}(a). Therefore one can conclude that the observed saturation magnetization decrease is also related to~crystallographic defects, such as APB, which arise in~the multi-variant growth, and whose persisent magnetic signal prevents the SRO film from its full saturation.

\begin{figure*}
	\center
	\includegraphics[scale=0.8]{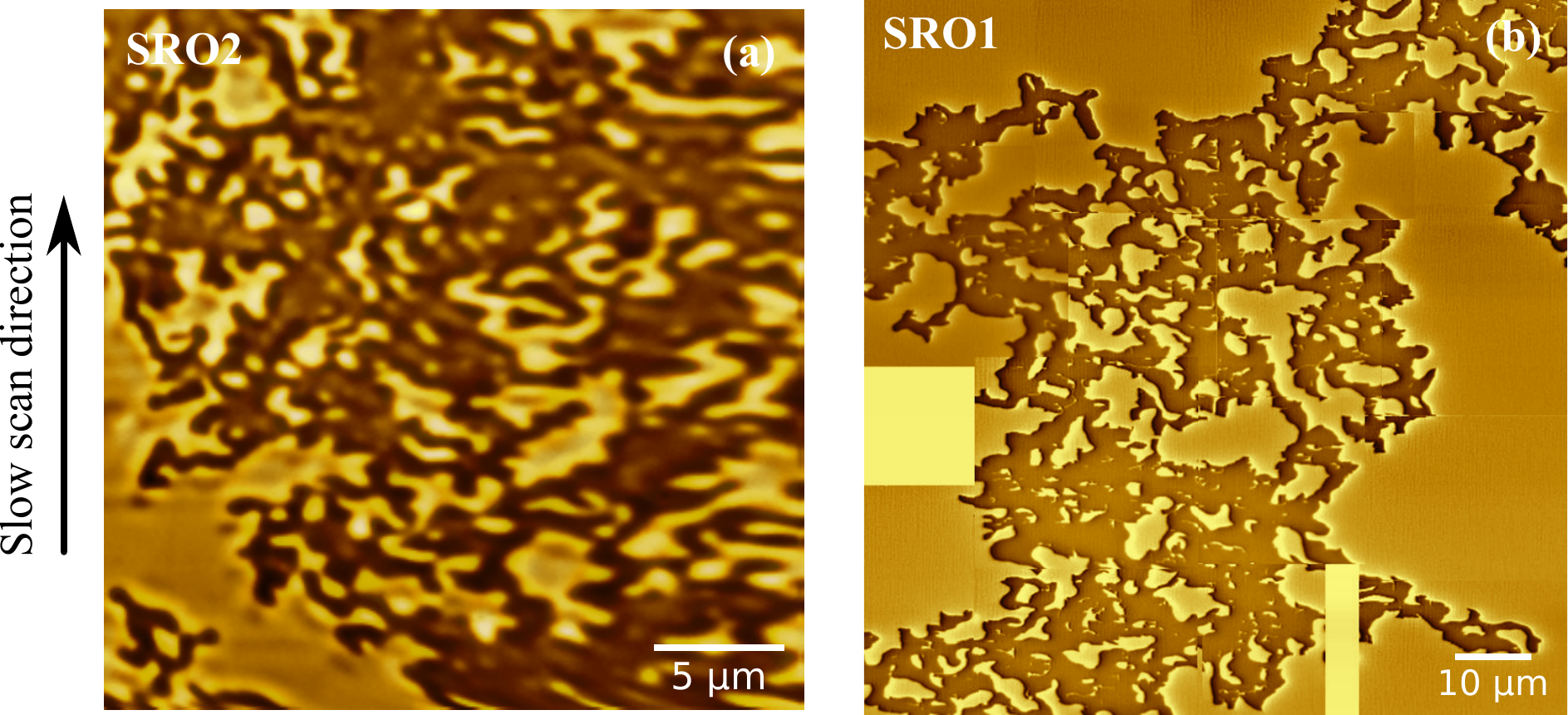}
	\caption{MFM images: (a) 26$\times$26~\si{\micro\metre\squared} area of~multi-variant SrRuO$_3$ film, measured at~20~K with field of~-180~mT applied perpendicular to~the sample surface. (b) 90$\times$90~\si{\micro\metre\squared} area in~single-variant SrRuO$_3$ film (SRO1) composed of~several 26$\times$26~\si{\micro\metre\squared} scans taken at~20~K, at~remanence after initiation of~the switching process with field of~-117~mT perpendicular to~the sample surface. Slow scan direction is indicated by~the black arrow.}
	\label{fig:SRO_MFM_large}
\end{figure*}

Fig.~\ref{fig:SRO_MFM_large}(a) shows a single scan of~SRO2 taken at~a higher negative field of~-180~mT. The scan was measured right after the field application, but here the aim is not to~discuss the dynamics. We want to~point out that more than a half of~the area is already reversed, which means that the overall magnetic moment of~the sample should be negative. However this scan was taken at~a field value determined as coercive field $\mu_0H_{C2,slow}$ according to~the SQUID magnetometry (see Fig.~\ref{fig:SQUID}(b)). This disagreement between MFM and SQUID points out the local character of~the MFM measurement, as only a small area of~the sample can be measured during the scan. Nevertheless, despite the quantitative inaccuracy of~the MFM, conclusions on~behaviour of~magnetic domains remain unequivocal. 

As demonstrated in~Fig.~\ref{fig:SRO2_MFM}, in~SRO2 sample we were able to~observe the domain nucleation process, which enabled determination of~the nucleation centers. On~the other hand this was not possible on~SRO1, where we did not succeed in~capturing the exact location of~the nucleation centers. As already presented earlier, in~Fig.~\ref{fig:SRO1_MFM}(b) the very beginning of~the switching process was captured, and yet a clearly demarcated area indicating a nucleation center is not visible. The reversed (dark) area is missing borders on~the left and top edge of~the image, suggesting that the domain nucleated outside the observed area. In~order to~locate the domain nucleation centers, we tried to~find one clearly demarcated reversed domain by~observing a larger area of~the sample. First a single MFM scan was measured, at~a small negative field of~-117~mT where a switching event appeared. Then the magnetic field was turned off to~prevent further propagation of~magnetic domains, and the scanner was moved around the area of~the sample to~find borders of~the initially observed reversed domain. This way an area of~90$\times$90 squared micrometers was investigated, as shown in~Fig.~\ref{fig:SRO_MFM_large}(b). However, we were not able to~find borders of~the reversed region, i.e. we did not locate the domain nucleation centers.

A remark should be made here, mentioning that one can also suggest an alternative explanation for the process depicted in~Fig.~\ref{fig:SRO1_MFM}. If one assumes that the domain nucleated in~the very center of~Fig.~\ref{fig:SRO1_MFM}(b), and that the reversal process then occured very quickly, it would be possible that the domain walls propagated from the center beyond the image borders before the scan was finished. In~order to~test this hypothesis, several control measurements were performed in~a following sequence. Firstly, the sample was saturated in~a high positive field, and a small negative field (e.g.~-120~mT) was applied to~initiate the nucleation process, after which a first image was captured. Secondly, the sample was saturated in~a high positive field again, and then a small negative field was applied, of~lower amplitude compared to~the first measurements (e.g.~-115~mT). Now a second image was captured and the domain pattern was compared with the first image. As the second image was recorded at~lower field, the reversed area should be smaller compared to~the first image, allowing to~determine the direction of~domain wall propagation, as well as the relative position of~the nucleation center. If~the nucleation center were located in~the middle of~the observed area, the domain walls would propagate from the middle, and the reversed area would be diminished at~the borders. However, the measurements revealed the exact opposite, i.e. the reversed area was significantly diminished in~the middle, suggesting that the domain wall propagation was proceeding from outside into the investigated area. This leaves the alternative hypothesis highly unlikely. The domain nucleation center always appeared to~be located outside the observed area. Together with the findings from Fig.~\ref{fig:SRO_MFM_large}(b), it leaves us unable to~determine the exact position of~domain nucleation centers.

Not being able to~capture the domain nucleation in~SRO1 indicates significantly lower density of~the nucleation centers in~SRO prepared on~vicinal STO substrate. Properties of~the magnetic domains are also different between the two films. SRO1 exhibits larger magnetic domains and smaller coercive field, which both indicate lower density of~pinning centers in~SRO on~vicinal substrate. The low density of~both the pinning and the nucleation centers is likely to~be related to~density of~crystallographic defects. Presence of~the defects is apparently suppressed in~SRO on~vicinal substrate via suppression of~the multi-variant growth. Even though the growth of~purely single-variant SRO film was not achieved, the representation of~second crystallographic variant in~case of~SRO1 is so low that the magnetic properties are notably improved. Absence of~the bubble-like features in~MFM measurements on~SRO1 further support their relation to~crystallographic defects, such as APB, which were reported as typical defects in~SRO thin films~\cite{Jiang1998a,Zakharov1999,Oh2007}. APB can lead to~antiferromagnetic ordering inducing magnetic signal that can persist up~to~high magnetic fields~\cite{Margulies1997}. Even though there are no MFM reports on~similar behaviour in~SRO films, a recent study reported almost identical MFM features arising near APB in~bulk Ni-Mn-Ga~\cite{Straka2018}. It can therefore be inferred that growth of~SRO on~vicinal STO substrate leads to~reduced density of~crystallographic defects acting as domain nucleation centers, such as APB, and consequently to significantly improved magnetic properties of~the films.

\section{Conclusions}

A study of~the influence of~substrate miscut on~magnetic properties of~SRO ultrathin films was performed. As expected, the structural investigation showed that multi-variant growth can be successfully suppressed by~use of~vicinal (high miscut angle) STO substrate. By~means of~SQUID magnetometry and MFM microscopy the magnetization dynamics and behaviour of~the magnetic domains was studied. Magnetization relaxation was found to~take place in~both the multi-variant and the nearly single-variant SRO films. The relaxation effects were observed on~similar timescale of~several hours in~both films. It was further found that the multi-variant film exhibits higher coercive field and smaller magnetic domains, which is directly related to~higher density of~pinning centers, i.e. higher density of~crystallographic defects. High density of~defects was confirmed also by~direct observation of~the domain nucleation centers, which are likely to~originate due to~the enhanced multi-variant growth. We believe that some of~the defects are anti-phase boundaries, leading to~antiferromagnetic ordering and persistent features in~MFM signal up~to~high magnetic fields. Presence of~such unsaturated magnetic structures results in~lower saturation magnetization of~the multi-variant film. Growth of~SRO on~vicinal STO substrate therefore leads to~reduced density of~crystallographic defects, i.e. to~better overall crystalline quality of~the films, and consequently to~improved magnetic properties of~SRO. Such results are of~high importance for design and further applications in~oxide spintronics and electronics, because it will allow direct tuning of~magnetic properties via substrate miscut angle or~other deposition parameters.

\section*{Acknowledgments}

The SQUID magnetometry and MFM microscopy experiments were performed in~MGML (www.mgml.eu), which is supported within the program of~Czech Research Infrastructures, Project No. LM2018096. The research was further supported by~the Ministry of~Education, Youth and Sports of~Czech Republic by~OP VVV, Project MATFUN No. \linebreak CZ.02.1.01/0.0/0.0/15\_003/0000487. This work was also supported by~GA\v{C}R, Project No. 19-09882S. This work was further supported by~a PHC Barrande grant of~the French Ministry for Europe and Foreign Affairs, Project No. 34000QK.

\section*{Data availability}

The raw data required to~reproduce these findings cannot be shared at~this time as the data also forms part of~an ongoing study. The processed data required to~reproduce these findings cannot be shared at~this time as the data also forms part of~an ongoing study.



\printcredits

\bibliographystyle{model1-num-names}

\bibliography{bibliografie}

\begin{thebibliography}{41}
\expandafter\ifx\csname natexlab\endcsname\relax\def\natexlab#1{#1}\fi
\providecommand{\url}[1]{\texttt{#1}}
\providecommand{\href}[2]{#2}
\providecommand{\path}[1]{#1}
\providecommand{\DOIprefix}{doi:}
\providecommand{\ArXivprefix}{arXiv:}
\providecommand{\URLprefix}{URL: }
\providecommand{\Pubmedprefix}{pmid:}
\providecommand{\doi}[1]{\href{http://dx.doi.org/#1}{\path{#1}}}
\providecommand{\Pubmed}[1]{\href{pmid:#1}{\path{#1}}}
\providecommand{\bibinfo}[2]{#2}
\ifx\xfnm\relax \def\xfnm[#1]{\unskip,\space#1}\fi
\bibitem[{Kanbayasi(1976)}]{Kanbayasi1976}
\bibinfo{author}{A.~Kanbayasi},
\newblock \bibinfo{title}{Magnetic properties of srruo3 single crystal},
\newblock \bibinfo{journal}{Journal of the Physical Society of Japan}
  \bibinfo{volume}{41} (\bibinfo{year}{1976}) \bibinfo{pages}{1876--1878}.
\bibitem[{Koster et~al.(2012)Koster, Klein, Siemons, Rijnders, Dodge, Eom,
  Blank, and Beasley}]{Koster2012}
\bibinfo{author}{G.~Koster}, \bibinfo{author}{L.~Klein},
  \bibinfo{author}{W.~Siemons}, \bibinfo{author}{G.~Rijnders},
  \bibinfo{author}{J.~S. Dodge}, \bibinfo{author}{C.-B. Eom},
  \bibinfo{author}{D.~H.~A. Blank}, \bibinfo{author}{M.~R. Beasley},
\newblock \bibinfo{title}{Structure, physical properties, and applications of
  ${\mathrm{srruo}}_{3}$ thin films},
\newblock \bibinfo{journal}{Rev. Mod. Phys.} \bibinfo{volume}{84}
  (\bibinfo{year}{2012}) \bibinfo{pages}{253--298}.
\bibitem[{Allouche et~al.(2016)Allouche, Gagou, Marrec, Fremy, and
  Marssi}]{Allouche2016}
\bibinfo{author}{B.~Allouche}, \bibinfo{author}{Y.~Gagou},
  \bibinfo{author}{F.~L. Marrec}, \bibinfo{author}{M.-A. Fremy},
  \bibinfo{author}{M.~E. Marssi},
\newblock \bibinfo{title}{Bipolar resistive switching and substrate effect in
  gdk2nb5o15 epitaxial thin films with tetragonal tungsten bronze type
  structure},
\newblock \bibinfo{journal}{Materials \& Design} \bibinfo{volume}{112}
  (\bibinfo{year}{2016}) \bibinfo{pages}{80 -- 87}.
\bibitem[{Herranz et~al.(2003)Herranz, Mart\'{i}nez, Fontcuberta, S\'{a}nchez,
  Garc\'{i}a-Cuenca, Ferrater, and Varela}]{Herranz2003}
\bibinfo{author}{G.~Herranz}, \bibinfo{author}{B.~Mart\'{i}nez},
  \bibinfo{author}{J.~Fontcuberta}, \bibinfo{author}{F.~S\'{a}nchez},
  \bibinfo{author}{M.~V. Garc\'{i}a-Cuenca}, \bibinfo{author}{C.~Ferrater},
  \bibinfo{author}{M.~Varela},
\newblock \bibinfo{title}{Srruo3/srtio3/srruo3 heterostructures for magnetic
  tunnel junctions},
\newblock \bibinfo{journal}{Journal of Applied Physics} \bibinfo{volume}{93}
  (\bibinfo{year}{2003}) \bibinfo{pages}{8035--8037}.
\bibitem[{Worledge and Geballe(2000)}]{Worledge2000}
\bibinfo{author}{D.~C. Worledge}, \bibinfo{author}{T.~H. Geballe},
\newblock \bibinfo{title}{Negative spin-polarization of
  ${\mathrm{srruo}}_{3}$},
\newblock \bibinfo{journal}{Phys. Rev. Lett.} \bibinfo{volume}{85}
  (\bibinfo{year}{2000}) \bibinfo{pages}{5182--5185}.
\bibitem[{Takahashi et~al.(2003)Takahashi, Sawa, Ishii, Akoh, Kawasaki, and
  Tokura}]{Takahashi2003}
\bibinfo{author}{K.~S. Takahashi}, \bibinfo{author}{A.~Sawa},
  \bibinfo{author}{Y.~Ishii}, \bibinfo{author}{H.~Akoh},
  \bibinfo{author}{M.~Kawasaki}, \bibinfo{author}{Y.~Tokura},
\newblock \bibinfo{title}{Inverse tunnel magnetoresistance in all-perovskite
  junctions of
  ${\mathrm{la}}_{0.7}{\mathrm{sr}}_{0.3}{\mathrm{mno}}_{3}/{\mathrm{srtio}}_{3}/{\mathrm{srruo}}_{3}$},
\newblock \bibinfo{journal}{Phys. Rev. B} \bibinfo{volume}{67}
  (\bibinfo{year}{2003}) \bibinfo{pages}{094413}.
\bibitem[{Ke et~al.(2004)Ke, Rzchowski, Belenky, and Eom}]{Ke2004}
\bibinfo{author}{X.~Ke}, \bibinfo{author}{M.~S. Rzchowski},
  \bibinfo{author}{L.~J. Belenky}, \bibinfo{author}{C.~B. Eom},
\newblock \bibinfo{title}{Positive exchange bias in ferromagnetic
  la0.67sr0.33mno3?srruo3 bilayers},
\newblock \bibinfo{journal}{Applied Physics Letters} \bibinfo{volume}{84}
  (\bibinfo{year}{2004}) \bibinfo{pages}{5458--5460}.
\bibitem[{Ke et~al.(2005)Ke, Belenky, Eom, and Rzchowski}]{Ke2005}
\bibinfo{author}{X.~Ke}, \bibinfo{author}{L.~J. Belenky},
  \bibinfo{author}{C.~B. Eom}, \bibinfo{author}{M.~S. Rzchowski},
\newblock \bibinfo{title}{Antiferromagnetic exchange-bias in epitaxial
  ferromagnetic la0.67sr0.33mno3?srruo3 bilayers},
\newblock \bibinfo{journal}{Journal of Applied Physics} \bibinfo{volume}{97}
  (\bibinfo{year}{2005}) \bibinfo{pages}{10K115}.
\bibitem[{Padhan et~al.(2006)Padhan, Prellier, and Budhani}]{Padhan2006}
\bibinfo{author}{P.~Padhan}, \bibinfo{author}{W.~Prellier},
  \bibinfo{author}{R.~C. Budhani},
\newblock \bibinfo{title}{Antiferromagnetic coupling and enhanced magnetization
  in all-ferromagnetic superlattices},
\newblock \bibinfo{journal}{Applied Physics Letters} \bibinfo{volume}{88}
  (\bibinfo{year}{2006}) \bibinfo{pages}{192509}.
\bibitem[{Solignac et~al.(2012)Solignac, Guerrero, Gogol, Maroutian, Ott,
  Largeau, Lecoeur, and Pannetier-Lecoeur}]{Solignac2012}
\bibinfo{author}{A.~Solignac}, \bibinfo{author}{R.~Guerrero},
  \bibinfo{author}{P.~Gogol}, \bibinfo{author}{T.~Maroutian},
  \bibinfo{author}{F.~Ott}, \bibinfo{author}{L.~Largeau},
  \bibinfo{author}{P.~Lecoeur}, \bibinfo{author}{M.~Pannetier-Lecoeur},
\newblock \bibinfo{title}{Dual antiferromagnetic coupling at
  ${\mathrm{la}}_{0.67}{\mathrm{sr}}_{0.33}{\mathrm{mno}}_{3}/{\mathrm{srruo}}_{3}$
  interfaces},
\newblock \bibinfo{journal}{Phys. Rev. Lett.} \bibinfo{volume}{109}
  (\bibinfo{year}{2012}) \bibinfo{pages}{027201}.
\bibitem[{Kurij(2016)}]{Kurij2016}
\bibinfo{author}{G.~Kurij}, \bibinfo{title}{Magnetic tunnel junctions for
  ultrasensitive all-oxide hybrid sensors for medical applications}, Ph.D.
  thesis, Universit\'{e} Paris-Saclay, \bibinfo{year}{2016}. \URLprefix
  \url{https://tel.archives-ouvertes.fr/tel-01359180}.
\bibitem[{Eom et~al.(1992)Eom, Cava, Fleming, Phillips, vanDover, Marshall,
  Hsu, Krajewski, and Peck}]{Eom1992}
\bibinfo{author}{C.~B. Eom}, \bibinfo{author}{R.~J. Cava},
  \bibinfo{author}{R.~M. Fleming}, \bibinfo{author}{J.~M. Phillips},
  \bibinfo{author}{R.~B. vanDover}, \bibinfo{author}{J.~H. Marshall},
  \bibinfo{author}{J.~W.~P. Hsu}, \bibinfo{author}{J.~J. Krajewski},
  \bibinfo{author}{W.~F. Peck},
\newblock \bibinfo{title}{Single-crystal epitaxial thin films of the isotropic
  metallic oxides sr1{\textendash}xcaxruo3 (0 <= x <= 1)},
\newblock \bibinfo{journal}{Science} \bibinfo{volume}{258}
  (\bibinfo{year}{1992}) \bibinfo{pages}{1766--1769}.
\bibitem[{Kats et~al.(2005)Kats, Genish, Klein, Reiner, and Beasley}]{Kats2005}
\bibinfo{author}{Y.~Kats}, \bibinfo{author}{I.~Genish},
  \bibinfo{author}{L.~Klein}, \bibinfo{author}{J.~W. Reiner},
  \bibinfo{author}{M.~R. Beasley},
\newblock \bibinfo{title}{Large anisotropy in the paramagnetic susceptibility
  of $\mathrm{Sr}\mathrm{Ru}{\mathrm{o}}_{3}$ films},
\newblock \bibinfo{journal}{Phys. Rev. B} \bibinfo{volume}{71}
  (\bibinfo{year}{2005}) \bibinfo{pages}{100403}.
\bibitem[{Klein et~al.(1996)Klein, Dodge, Ahn, Reiner, Mieville, Geballe,
  Beasley, and Kapitulnik}]{Klein1996}
\bibinfo{author}{L.~Klein}, \bibinfo{author}{J.~S. Dodge},
  \bibinfo{author}{C.~H. Ahn}, \bibinfo{author}{J.~W. Reiner},
  \bibinfo{author}{L.~Mieville}, \bibinfo{author}{T.~H. Geballe},
  \bibinfo{author}{M.~R. Beasley}, \bibinfo{author}{A.~Kapitulnik},
\newblock \bibinfo{title}{Transport and magnetization in the badly metallic
  itinerant ferromagnet srruo$_3$},
\newblock \bibinfo{journal}{Journal of Physics: Condensed Matter}
  \bibinfo{volume}{8} (\bibinfo{year}{1996}) \bibinfo{pages}{10111}.
\bibitem[{Jiang et~al.(1998)Jiang, Tian, Pan, Gan, and Eom}]{Jiang1998a}
\bibinfo{author}{J.~C. Jiang}, \bibinfo{author}{W.~Tian},
  \bibinfo{author}{X.~Q. Pan}, \bibinfo{author}{Q.~Gan}, \bibinfo{author}{C.~B.
  Eom},
\newblock \bibinfo{title}{Domain structure of epitaxial srruo3 thin films on
  miscut (001) srtio3 substrates},
\newblock \bibinfo{journal}{Applied Physics Letters} \bibinfo{volume}{72}
  (\bibinfo{year}{1998}) \bibinfo{pages}{2963--2965}.
\bibitem[{Marshall et~al.(1999)Marshall, Klein, Dodge, Ahn, Reiner, Mieville,
  Antagonazza, Kapitulnik, Geballe, and Beasley}]{Marshall1999}
\bibinfo{author}{A.~F. Marshall}, \bibinfo{author}{L.~Klein},
  \bibinfo{author}{J.~S. Dodge}, \bibinfo{author}{C.~H. Ahn},
  \bibinfo{author}{J.~W. Reiner}, \bibinfo{author}{L.~Mieville},
  \bibinfo{author}{L.~Antagonazza}, \bibinfo{author}{A.~Kapitulnik},
  \bibinfo{author}{T.~H. Geballe}, \bibinfo{author}{M.~R. Beasley},
\newblock \bibinfo{title}{Lorentz transmission electron microscope study of
  ferromagnetic domain walls in srruo3: Statics, dynamics, and crystal
  structure correlation},
\newblock \bibinfo{journal}{Journal of Applied Physics} \bibinfo{volume}{85}
  (\bibinfo{year}{1999}) \bibinfo{pages}{4131--4140}.
\bibitem[{Gan et~al.(1999)Gan, Rao, Eom, Wu, and Tsui}]{Gan1999}
\bibinfo{author}{Q.~Gan}, \bibinfo{author}{R.~A. Rao}, \bibinfo{author}{C.~B.
  Eom}, \bibinfo{author}{L.~Wu}, \bibinfo{author}{F.~Tsui},
\newblock \bibinfo{title}{Lattice distortion and uniaxial magnetic anisotropy
  in single domain epitaxial (110) films of srruo3},
\newblock \bibinfo{journal}{Journal of Applied Physics} \bibinfo{volume}{85}
  (\bibinfo{year}{1999}) \bibinfo{pages}{5297--5299}.
\bibitem[{Kolesnik et~al.(2006)Kolesnik, Yoo, Chmaissem, Dabrowski, Maxwell,
  Kimball, and Genis}]{Kolesnik2006}
\bibinfo{author}{S.~Kolesnik}, \bibinfo{author}{Y.~Z. Yoo},
  \bibinfo{author}{O.~Chmaissem}, \bibinfo{author}{B.~Dabrowski},
  \bibinfo{author}{T.~Maxwell}, \bibinfo{author}{C.~W. Kimball},
  \bibinfo{author}{A.~P. Genis},
\newblock \bibinfo{title}{Effect of crystalline quality and substitution on
  magnetic anisotropy of srruo3 thin films},
\newblock \bibinfo{journal}{Journal of Applied Physics} \bibinfo{volume}{99}
  (\bibinfo{year}{2006}) \bibinfo{pages}{08F501}.
\bibitem[{Feigenson et~al.(2008)Feigenson, Reiner, and Klein}]{Feigenson2008}
\bibinfo{author}{M.~Feigenson}, \bibinfo{author}{J.~W. Reiner},
  \bibinfo{author}{L.~Klein},
\newblock \bibinfo{title}{Current-induced magnetic instability in srruo3},
\newblock \bibinfo{journal}{Journal of Applied Physics} \bibinfo{volume}{103}
  (\bibinfo{year}{2008}) \bibinfo{pages}{07E741}.
\bibitem[{Feigenson et~al.(2007)Feigenson, Reiner, and Klein}]{Feigenson2007}
\bibinfo{author}{M.~Feigenson}, \bibinfo{author}{J.~W. Reiner},
  \bibinfo{author}{L.~Klein},
\newblock \bibinfo{title}{Efficient current-induced domain-wall displacement in
  ${\mathrm{srruo}}_{3}$},
\newblock \bibinfo{journal}{Phys. Rev. Lett.} \bibinfo{volume}{98}
  (\bibinfo{year}{2007}) \bibinfo{pages}{247204}.
\bibitem[{Sarkar et~al.(2013)Sarkar, Dalal, and De}]{Sarkar2013}
\bibinfo{author}{B.~Sarkar}, \bibinfo{author}{B.~Dalal}, \bibinfo{author}{S.~K.
  De},
\newblock \bibinfo{title}{Temperature induced magnetization reversal in
  srruo3},
\newblock \bibinfo{journal}{Applied Physics Letters} \bibinfo{volume}{103}
  (\bibinfo{year}{2013}) \bibinfo{pages}{252403}.
\bibitem[{Shperber et~al.(2012)Shperber, Bedau, Reiner, and
  Klein}]{Shperber2012}
\bibinfo{author}{Y.~Shperber}, \bibinfo{author}{D.~Bedau},
  \bibinfo{author}{J.~W. Reiner}, \bibinfo{author}{L.~Klein},
\newblock \bibinfo{title}{Current-induced magnetization reversal in
  srruo${}_{3}$},
\newblock \bibinfo{journal}{Phys. Rev. B} \bibinfo{volume}{86}
  (\bibinfo{year}{2012}) \bibinfo{pages}{085102}.
\bibitem[{Shperber et~al.(2013)Shperber, Sinwani, Naftalis, Bedau, Reiner, and
  Klein}]{Shperber2013}
\bibinfo{author}{Y.~Shperber}, \bibinfo{author}{O.~Sinwani},
  \bibinfo{author}{N.~Naftalis}, \bibinfo{author}{D.~Bedau},
  \bibinfo{author}{J.~W. Reiner}, \bibinfo{author}{L.~Klein},
\newblock \bibinfo{title}{Thermally assisted current-induced magnetization
  reversal in srruo${}_{3}$},
\newblock \bibinfo{journal}{Phys. Rev. B} \bibinfo{volume}{87}
  (\bibinfo{year}{2013}) \bibinfo{pages}{115118}.
\bibitem[{Zhou et~al.(2014)Zhou, Li, Xiong, Zhang, Wang, Cao, Lv, and
  Du}]{Zhou2014}
\bibinfo{author}{W.~P. Zhou}, \bibinfo{author}{Q.~Li}, \bibinfo{author}{Y.~Q.
  Xiong}, \bibinfo{author}{Q.~M. Zhang}, \bibinfo{author}{D.~H. Wang},
  \bibinfo{author}{Q.~Q. Cao}, \bibinfo{author}{L.~Y. Lv},
  \bibinfo{author}{Y.~W. Du},
\newblock \bibinfo{title}{Electric field manipulation of~magnetic and transport
  properties in~srruo3/pb(mg1/3nb2/3)o3-pbtio3 heterostructure},
\newblock \bibinfo{journal}{Scientific Reports}  (\bibinfo{year}{2014}).
\bibitem[{Barkhausen(1919)}]{Barkhausen1919}
\bibinfo{author}{H.~Barkhausen},
\newblock \bibinfo{title}{Zwei mit hilfe der neuen verst{\"a}rker entdeckte
  erscheinungen},
\newblock \bibinfo{journal}{Phys. Z} \bibinfo{volume}{20}
  (\bibinfo{year}{1919}) \bibinfo{pages}{401}.
\bibitem[{Pommier et~al.(1990)Pommier, Meyer, P\'enissard, Ferr\'e, Bruno, and
  Renard}]{Pommier1990}
\bibinfo{author}{J.~Pommier}, \bibinfo{author}{P.~Meyer},
  \bibinfo{author}{G.~P\'enissard}, \bibinfo{author}{J.~Ferr\'e},
  \bibinfo{author}{P.~Bruno}, \bibinfo{author}{D.~Renard},
\newblock \bibinfo{title}{Magnetization reversal in ultrathin ferromagnetic
  films with perpendicular anistropy: Domain observations},
\newblock \bibinfo{journal}{Phys. Rev. Lett.} \bibinfo{volume}{65}
  (\bibinfo{year}{1990}) \bibinfo{pages}{2054--2057}.
\bibitem[{Schwarz et~al.(2004)Schwarz, Liebmann, Kaiser, Wiesendanger, Noh, and
  Kim}]{Schwarz2004}
\bibinfo{author}{A.~Schwarz}, \bibinfo{author}{M.~Liebmann},
  \bibinfo{author}{U.~Kaiser}, \bibinfo{author}{R.~Wiesendanger},
  \bibinfo{author}{T.~W. Noh}, \bibinfo{author}{D.~W. Kim},
\newblock \bibinfo{title}{Visualization of the barkhausen effect by magnetic
  force microscopy},
\newblock \bibinfo{journal}{Phys. Rev. Lett.} \bibinfo{volume}{92}
  (\bibinfo{year}{2004}) \bibinfo{pages}{077206}.
\bibitem[{Landau et~al.(2012)Landau, Reiner, and Klein}]{Landau2012}
\bibinfo{author}{L.~Landau}, \bibinfo{author}{J.~W. Reiner},
  \bibinfo{author}{L.~Klein},
\newblock \bibinfo{title}{Low temperature magnetic force microscope study of
  magnetization reversal in patterned nanoislands of srruo3},
\newblock \bibinfo{journal}{Journal of Applied Physics} \bibinfo{volume}{111}
  (\bibinfo{year}{2012}) \bibinfo{pages}{07B901}.
\bibitem[{Jiang et~al.(1998)Jiang, Tian, Pan, Gan, and Eom}]{Jiang1998b}
\bibinfo{author}{J.~Jiang}, \bibinfo{author}{W.~Tian},
  \bibinfo{author}{X.~Pan}, \bibinfo{author}{Q.~Gan}, \bibinfo{author}{C.~Eom},
\newblock \bibinfo{title}{Effects of miscut of the {SrTiO$_3$} substrate on
  microstructures of the epitaxial {SrRuO$_3$} thin films},
\newblock \bibinfo{journal}{Materials Science and Engineering: B}
  \bibinfo{volume}{56} (\bibinfo{year}{1998}) \bibinfo{pages}{152 -- 157}.
\bibitem[{Gan et~al.(1997)Gan, Rao, and Eom}]{Gan1997}
\bibinfo{author}{Q.~Gan}, \bibinfo{author}{R.~A. Rao}, \bibinfo{author}{C.~B.
  Eom},
\newblock \bibinfo{title}{Control of the growth and domain structure of
  epitaxial {SrRuO$_3$} thin films by vicinal (001) {SrTiO$_3$} substrates},
\newblock \bibinfo{journal}{Applied Physics Letters} \bibinfo{volume}{70}
  (\bibinfo{year}{1997}) \bibinfo{pages}{1962--1964}.
\bibitem[{Ne\v{c}as and Klapetek(2012)}]{Necas2012}
\bibinfo{author}{D.~Ne\v{c}as}, \bibinfo{author}{P.~Klapetek},
\newblock \bibinfo{title}{Gwyddion: an open-source software for {SPM} data
  analysis},
\newblock \bibinfo{journal}{Central European Journal of Physics}
  \bibinfo{volume}{10} (\bibinfo{year}{2012}) \bibinfo{pages}{181--188}.
\bibitem[{Vailionis et~al.(2007)Vailionis, Siemons, and Koster}]{Vailionis2007}
\bibinfo{author}{A.~Vailionis}, \bibinfo{author}{W.~Siemons},
  \bibinfo{author}{G.~Koster},
\newblock \bibinfo{title}{Strain-induced single-domain growth of epitaxial
  srruo$_3$ layers on srtio$_3$: A high-temperature x-ray diffraction study},
\newblock \bibinfo{journal}{Applied Physics Letters} \bibinfo{volume}{91}
  (\bibinfo{year}{2007}) \bibinfo{pages}{071907}.
\bibitem[{Est\`eve et~al.(2011)Est\`eve, Maroutian, Pillard, and
  Lecoeur}]{Esteve2011}
\bibinfo{author}{D.~Est\`eve}, \bibinfo{author}{T.~Maroutian},
  \bibinfo{author}{V.~Pillard}, \bibinfo{author}{P.~Lecoeur},
\newblock \bibinfo{title}{Step velocity tuning of {SrRuO${}_{3}$} step flow
  growth on {SrTiO${}_{3}$}},
\newblock \bibinfo{journal}{Phys. Rev. B} \bibinfo{volume}{83}
  (\bibinfo{year}{2011}) \bibinfo{pages}{193401}.
\bibitem[{Bushmeleva et~al.(2006)Bushmeleva, Pomjakushin, Pomjakushina,
  Sheptyakov, and Balagurov}]{Bushmeleva2006}
\bibinfo{author}{S.~Bushmeleva}, \bibinfo{author}{V.~Pomjakushin},
  \bibinfo{author}{E.~Pomjakushina}, \bibinfo{author}{D.~Sheptyakov},
  \bibinfo{author}{A.~Balagurov},
\newblock \bibinfo{title}{Evidence for the band ferromagnetism in {SrRuO$_3$}
  from neutron diffraction},
\newblock \bibinfo{journal}{Journal of Magnetism and Magnetic Materials}
  \bibinfo{volume}{305} (\bibinfo{year}{2006}) \bibinfo{pages}{491 -- 496}.
\bibitem[{Xia et~al.(2009)Xia, Siemons, Koster, Beasley, and
  Kapitulnik}]{Xia2009}
\bibinfo{author}{J.~Xia}, \bibinfo{author}{W.~Siemons},
  \bibinfo{author}{G.~Koster}, \bibinfo{author}{M.~R. Beasley},
  \bibinfo{author}{A.~Kapitulnik},
\newblock \bibinfo{title}{Critical thickness for itinerant ferromagnetism in
  ultrathin films of ${\text{srruo}}_{3}$},
\newblock \bibinfo{journal}{Phys. Rev. B} \bibinfo{volume}{79}
  (\bibinfo{year}{2009}) \bibinfo{pages}{140407}.
\bibitem[{Ishigami et~al.(2015)Ishigami, Yoshimatsu, Toyota, Takizawa, Yoshida,
  Shibata, Harano, Takahashi, Kadono, Verma, Singh, Takeda, Okane, Saitoh,
  Yamagami, Koide, Oshima, Kumigashira, and Fujimori}]{Ishigami2015}
\bibinfo{author}{K.~Ishigami}, \bibinfo{author}{K.~Yoshimatsu},
  \bibinfo{author}{D.~Toyota}, \bibinfo{author}{M.~Takizawa},
  \bibinfo{author}{T.~Yoshida}, \bibinfo{author}{G.~Shibata},
  \bibinfo{author}{T.~Harano}, \bibinfo{author}{Y.~Takahashi},
  \bibinfo{author}{T.~Kadono}, \bibinfo{author}{V.~K. Verma},
  \bibinfo{author}{V.~R. Singh}, \bibinfo{author}{Y.~Takeda},
  \bibinfo{author}{T.~Okane}, \bibinfo{author}{Y.~Saitoh},
  \bibinfo{author}{H.~Yamagami}, \bibinfo{author}{T.~Koide},
  \bibinfo{author}{M.~Oshima}, \bibinfo{author}{H.~Kumigashira},
  \bibinfo{author}{A.~Fujimori},
\newblock \bibinfo{title}{Thickness-dependent magnetic properties and
  strain-induced orbital magnetic moment in ${\mathrm{srruo}}_{3}$ thin films},
\newblock \bibinfo{journal}{Phys. Rev. B} \bibinfo{volume}{92}
  (\bibinfo{year}{2015}) \bibinfo{pages}{064402}.
\bibitem[{Zijlstra(1979)}]{Zijlstra1979}
\bibinfo{author}{H.~Zijlstra},
\newblock \bibinfo{title}{Coping with brown's paradox: The pinning and
  nucleation of magnetic domain walls at antiphase boundaries},
\newblock \bibinfo{journal}{IEEE Transactions on Magnetics}
  \bibinfo{volume}{15} (\bibinfo{year}{1979}) \bibinfo{pages}{1246--1250}.
\bibitem[{Margulies et~al.(1997)Margulies, Parker, Rudee, Spada, Chapman,
  Aitchison, and Berkowitz}]{Margulies1997}
\bibinfo{author}{D.~T. Margulies}, \bibinfo{author}{F.~T. Parker},
  \bibinfo{author}{M.~L. Rudee}, \bibinfo{author}{F.~E. Spada},
  \bibinfo{author}{J.~N. Chapman}, \bibinfo{author}{P.~R. Aitchison},
  \bibinfo{author}{A.~E. Berkowitz},
\newblock \bibinfo{title}{{O}rigin of the {A}nomalous {M}agnetic {B}ehavior in
  {S}ingle {C}rystal {${\mathrm{Fe}}_{3}{O}_{4}$} {F}ilms},
\newblock \bibinfo{journal}{Phys. Rev. Lett.} \bibinfo{volume}{79}
  (\bibinfo{year}{1997}) \bibinfo{pages}{5162--5165}.
\bibitem[{Zakharov et~al.(1999)Zakharov, Satyalakshmi, Koren, and
  Hesse}]{Zakharov1999}
\bibinfo{author}{N.~D. Zakharov}, \bibinfo{author}{K.~M. Satyalakshmi},
  \bibinfo{author}{G.~Koren}, \bibinfo{author}{D.~Hesse},
\newblock \bibinfo{title}{Substrate temperature dependence of structure and
  resistivity of {SrRuO$_3$} thin films grown by pulsed laser deposition on
  (100) {SrTiO$_3$}},
\newblock \bibinfo{journal}{Journal of Materials Research} \bibinfo{volume}{14}
  (\bibinfo{year}{1999}) \bibinfo{pages}{4385--4394}.
\bibitem[{Oh et~al.(2007)Oh, Suh, and Park}]{Oh2007}
\bibinfo{author}{S.~H. Oh}, \bibinfo{author}{J.~H. Suh}, \bibinfo{author}{C.~G.
  Park},
\newblock \bibinfo{title}{{D}efects in {S}trained {E}pitaxial {SrRuO$_3$}
  {F}ilms on {SrTiO$_3$} {S}ubstrates},
\newblock \bibinfo{journal}{MATERIALS TRANSACTIONS} \bibinfo{volume}{48}
  (\bibinfo{year}{2007}) \bibinfo{pages}{2556--2562}.
\bibitem[{Straka et~al.(2018)Straka, Fekete, and Heczko}]{Straka2018}
\bibinfo{author}{L.~Straka}, \bibinfo{author}{L.~Fekete},
  \bibinfo{author}{O.~Heczko},
\newblock \bibinfo{title}{Antiphase boundaries in bulk {Ni-Mn-Ga} {H}eusler
  alloy observed by magnetic force microscopy},
\newblock \bibinfo{journal}{Applied Physics Letters} \bibinfo{volume}{113}
  (\bibinfo{year}{2018}) \bibinfo{pages}{172901}.

\end{thebibliography}


%
%

\end{document}